\documentclass[prb,a4paper,showpacs,showkeys,superscriptaddress]{revtex4}

\usepackage{latexsym}
\usepackage{amsmath}
\usepackage{bm}
\usepackage{dcolumn}

\usepackage{color}

\usepackage[dvips]{graphicx}
\graphicspath{{./}}

\usepackage[latin1]{inputenc}
\usepackage{subfigure}

\newcommand {\qref}[1]{Ref.~\onlinecite{#1}}

\newcommand {\qfig}[1]{Fig.~\ref{f_#1}}
\newcommand {\queq}[1]{(\ref{#1})}
\newcommand {\qeq}[1]{Eq.~\queq{#1}}

\newcommand {\Om}{\bm{\Omega}}

\newcommand {\Eriso}{M^{\rm iso}}
\newcommand {\Er}{M}
\newcommand {\ErOm}{M(\Om)}

\newcommand {\Erfitshort}{M^{\rm MD, init}}
\newcommand {\Erfitlong}{M^{\rm MD, full}}

\newcommand {\Erapp}{M^{\rm appr}}
\newcommand {\Ertheo}{M^{\rm theo}}

\newcommand {\Aatom}{A^{\rm atom}}
\newcommand {\Apro}{A^{\rm proj}}
\newcommand {\xmax}{x^{\rm max}}
\newcommand {\xmin}{x^{\rm min}}
\newcommand {\ymax}{y^{\rm max}}
\newcommand {\ymin}{y^{\rm min}}

\newcommand {\eVA}{eV/\AA$^3$}
\newcommand {\Dt}{\Delta t}
\newcommand {\ani}{X}

\newcommand {\slip}{$\langle 1\bar{1}0\rangle$ \{111\}}
\newcommand {\szero}{s_{100}}
\newcommand {\sone}{s_{111}}

\newcommand {\rcut}{r_{\rm cut}}

\begin{document}

\date{\today}

\title{Influence of crystal anisotropy on elastic deformation and onset
of plasticity in nanoindentation -- a simulational study}

\author{Gerolf Ziegenhain}
\affiliation{%
Fachbereich Physik und Forschungszentrum OPTIMAS,
Universit{\"a}t Kaiserslautern, \\
Erwin-Schr{\"o}dinger-Stra{\ss}e, D-67663 Kaiserslautern, Germany}

 \author{Herbert M.~Urbassek}
 \email{urbassek@rhrk.uni-kl.de}
 \homepage{ http://www.physik.uni-kl.de/urbassek/}
\affiliation{%
Fachbereich Physik und Forschungszentrum OPTIMAS,
Universit{\"a}t Kaiserslautern, \\
Erwin-Schr{\"o}dinger-Stra{\ss}e, D-67663 Kaiserslautern, Germany}

\author{Alexander Hartmaier}
\affiliation{%
Interdisciplinary Center for Advanced Materials Simulation (ICAMS),
Ruhr-University Bochum,\\ Stiepeler Str.\ 129 (UHW), D-44780 Bochum,
Germany}

\begin{abstract}

Using molecular-dynamics simulation we
simulate nanoindentation into the three principal surfaces -- the (100),
(110) and (111) surface -- of Cu and Al. In the elastic regime, the
simulation data agree fairly well with the linear elastic theory of
indentation into an elastically anisotropic substrate. With increasing
indentation, the effect of pressure hardening becomes visible. When the
critical stress for dislocation nucleation is reached, even the
elastically isotropic Al shows a strong dependence of the surface
orientation on the force-displacement curves. After the load drop, when
plasticity has set in, the influence of the surface orientation is
lost, and the contact pressure (hardness) becomes independent of the
surface orientation.

\end{abstract}



\maketitle


\section{Introduction}

In a seminal paper, Hertz investigated the elastic interaction between a
sphere of radius $R$ with an elastic isotropic solid.\cite{Her82} He
found that the force $F$ acting on the sphere normally to the surface
and the displacement $d$ into the surface are related by the so-called
Hertz law

\begin{equation} \label{Hertz}
 F = \frac43 \Er d^{3/2} \sqrt{R} .
\end{equation}

In this relation, a single materials parameter, the so-called
indentation modulus
$\Er$ describes the materials elastic
response.\cite{Her82,LLvii,Fis04,Fis07} For
a rigid indenter, it may be expressed in terms of the Young's modulus
$E$ and the Poisson ratio $\nu$ of the substrate as

\begin{equation} \label{Er}
\Er = \frac{E}{1-\nu^2} .
\end{equation}

Hertz also determined the contact pressure; it is defined by the ratio
of the normal force $F$  divided by the contact area projected into the
surface plane. Hertz obtained

\begin{equation} \label{p}
p = \frac{4}{3\pi} \Er \sqrt{ \frac{d}{R} } .
\end{equation}

We note that Hertz' analysis is exact in the limit of $d \ll R$.

Nowadays indentation experiments can be performed into single
crystals.\cite{KH98,KJHR99} The question then arises in how far Hertz'
analysis also describes the indentation of crystalline -- and thus by
definition anisotropic -- materials. An analytical extension of Hertz'
analysis to anisotropic materials is, however, non-trivial.
Willis\cite{Wil66a,Wil67} appears to have gone farthest in the analysis;
however, no analytical results are available, and one has to resort to
numerical procedures even in the simplest case, i.e., for transversally
isotropic media. Vlassak and Nix\cite{VN93,VN94} evaluated these results
numerically for the specific problems of a flat circular punch and an
axisymmetric paraboloid indenter; the latter approximates a spherical
indenter for small indentation depths. They showed that in this case the
Hertzian law \qeq{Hertz} holds with a modified indentation modulus
$\ErOm$ which depends on the surface orientation $\Om$, and calculated
this quantity numerically. In their experiments on Au single crystals,
Kiely and Houston found considerable deviations between the experimental
results on a Au substrate and the theoretical predictions.\cite{KH98}
Recently, Tsuru and Shibutani employed molecular-dynamics (MD)
simulations to study indentation into an anisotropic half-space,
and demonstrated that indentation into fcc crystals indeed
depends on the surface orientation.\cite{TS06,TS07}  Other authors also
investigated the question in how far the homogeneous nucleation of
dislocations under the indenter, i.e., the onset of plasticity, is
influenced by the surface orientation.\cite{VLZ*03,TSM07,LGSC08}

In the present paper, we study these questions for two fcc materials, Al
and Cu, which show a widely differing anisotropy. This allows us to
discuss quantitatively in how far the Hertzian law, \queq{Hertz}, is
obeyed. Finally, we demonstrate that with increasing penetration into
the target, the indenter measures the bulk hardness of the system and
the surface orientation loses its influence.

\section{Method}

\subsection{Simulation}

We chose two different fcc materials, Al and Cu, for our study, which
differ strongly in their elastic anisotropy. Many-body potentials of the
embedded-atom type are used to model these metals; both the Al
potential\cite{ZWJ*01} and the Cu potential\cite{MMP*01} reproduce the
zero-temperature elastic moduli given in Table I. The fcc substrate has
approximately cubic shape with side lengths of around 25 nm; it
contains roughly $1.35\times 10^6$ atoms. We checked in a series of
simulations that our crystallite size is large enough to obtain reliable
results for the indentation process. We found that a careful relaxation
of the crystal before starting the indentation process to $p_{ij} <
10^{-5}$ GPa and temperatures $\ll 1$ K was mandatory to obtain reliable
and reproducible results. Lateral periodic boundary conditions have been
applied. At the bottom, atoms in a layer of the width $\rcut$ have been
constrained to $F_{\rm normal} = 0$.

The indenter is modelled as a repulsive sphere. We chose a non-atomistic
representation of the indenter. Its interaction potential with the
substrate atoms is described by\cite{KPH98}

\begin{equation} \label{5}
V(r) = \left\{ \begin{array}{ll}
k(R-r)^3, & r<R, \\
0, & r\ge R .
\end{array} \right.
\end{equation}

The indenter radius was set to $R=8$ nm, and the indenter stiffness to
$k=3$ \eVA. We checked that our results are only weakly influenced by
the exact value of the contact stiffness, as long as it is in the range
of $1-10$ \eVA.

The simulations have been performed using a modified version of the
LAMMPS code,\cite{LAMMPS} using the so-called displacement-controlled
approach.\cite{CSR01,MY03} The indenter is advanced every $\Dt=2$ ps by
a fixed amount of $\delta= 0.256$ \AA\ ($\ll$ lattice constant)
instantaneously, corresponding to an average indentation speed of
$v=12.8$ m/s. The substrate then relaxes for the ensuing time of $\Dt$
to the new indenter position.

\subsection{Elastic properties}

\begin{figure}

\subfigure[]{\includegraphics[width=0.45\textwidth]{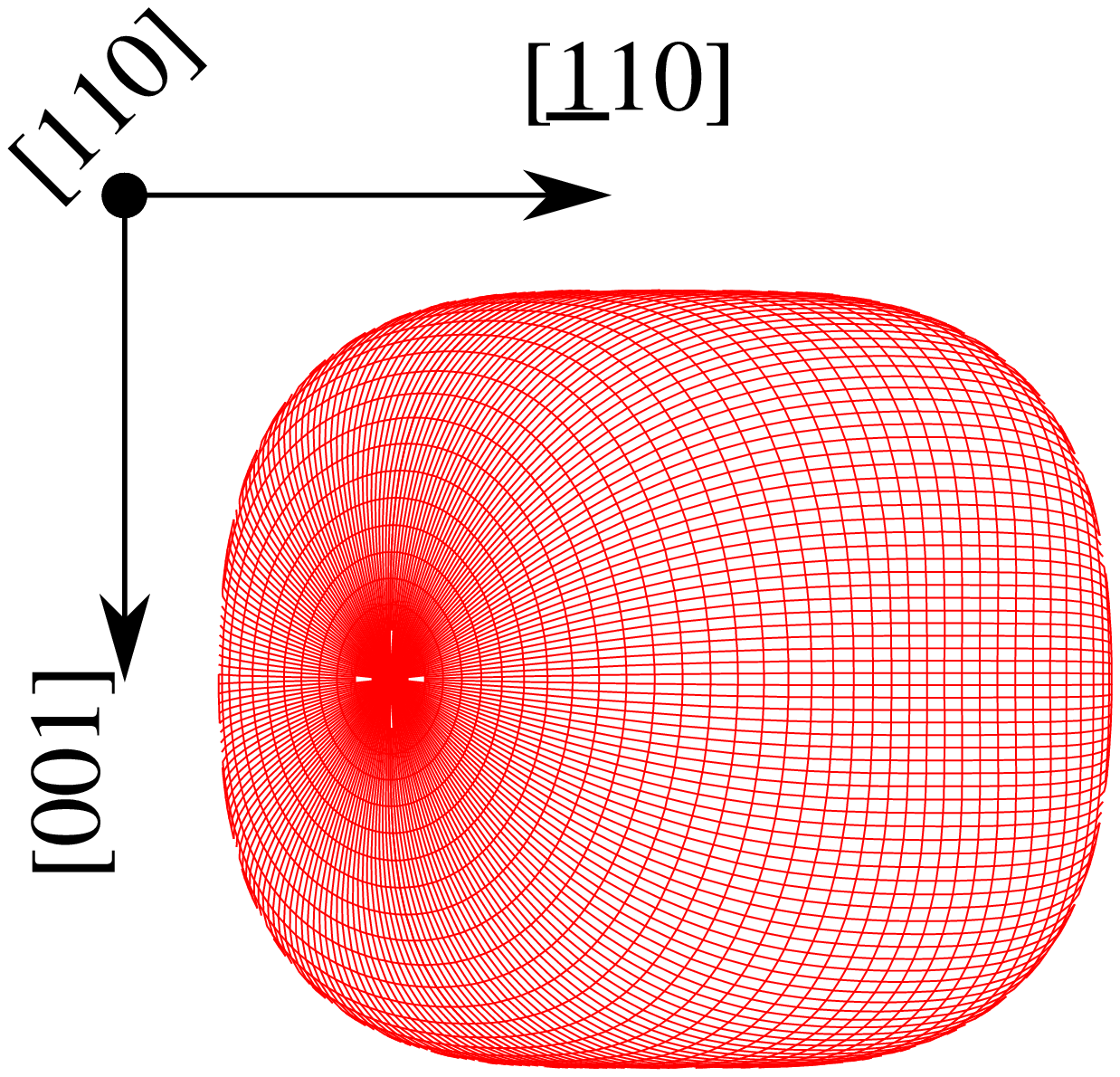}}
\hfill
\subfigure[]{\includegraphics[width=0.45\textwidth]{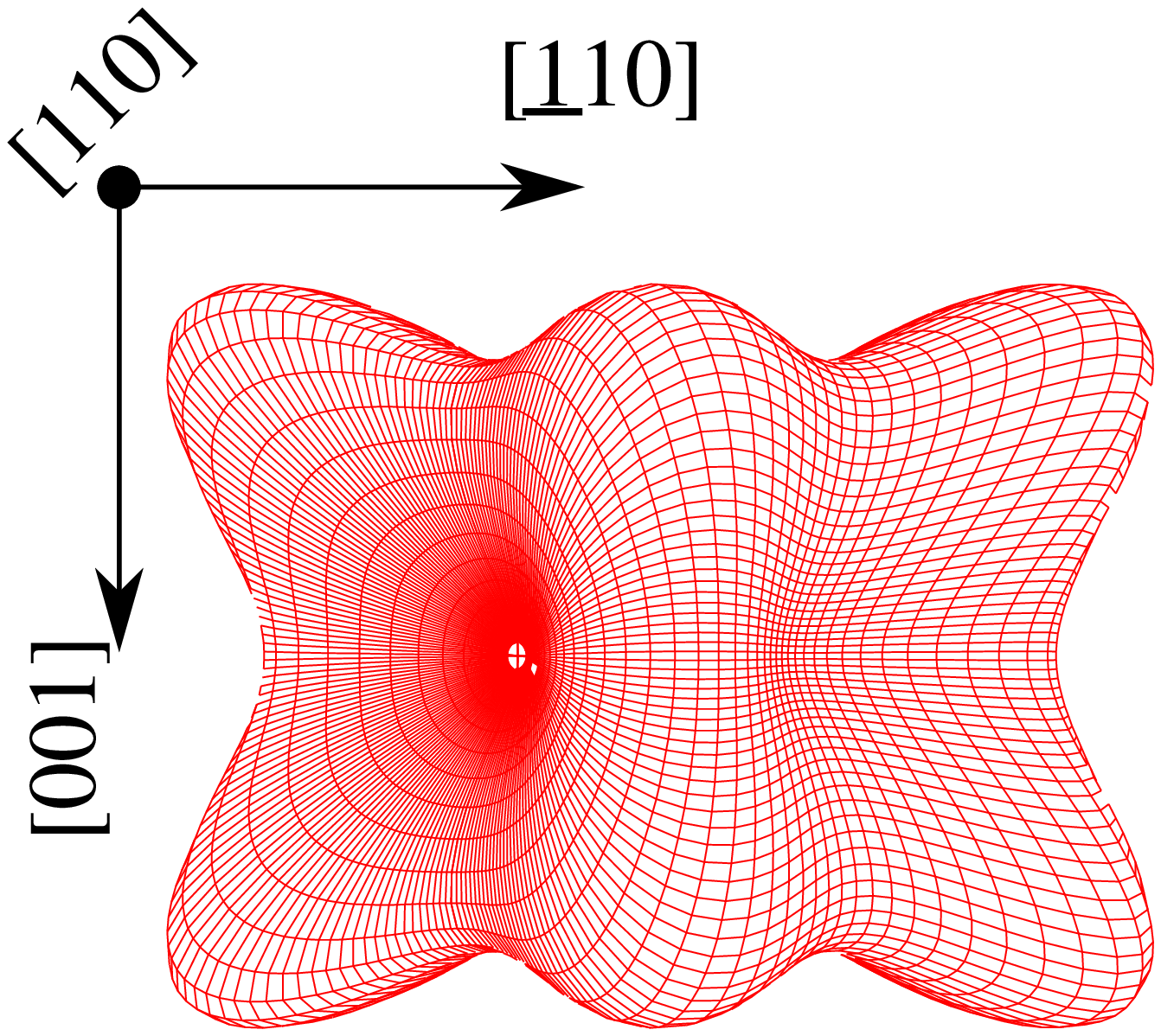}}

\caption{ Polar plot of the orientation dependence of Young's modulus,
for Al (a) and Cu (b), \qeq{EOm}.  The view direction is aligned with the
[110] direction.
 }

\label{f_aniso}
\end{figure}

The materials considered here crystallize in the cubic fcc structure.
Their elastic behaviour is therefore completely described by their
elastic constants, $c_{11}$, $c_{12}$, and $c_{44}$. Their
zero-temperature values,
are given in Table I. These two materials differ in particular
in their elastic anisotropy, defined as

\begin{equation} \label{ani}
\ani = \frac{2c_{44}}{c_{11} - c_{12}} ,
\end{equation}

which measures both the orientation dependence of the elastic modulus
and of the shear moduli. As Table I shows, Cu is quite anisotropic,
$\ani=3.22$, while Al is nearly isotropic, $\ani=1.20$. \qfig{aniso}
demonstrates the anisotropy of the two materials by plotting
Young's modulus, $E(\Om)$, in its dependence on orientation $\Om$.  It
is given by

\begin{equation} \label{EOm}
\frac1{E(\Om)} = \frac1E -\frac1{c_{44}} (\ani-1) \Gamma(\Om),
\end{equation}

where $E$ is the orientation-averaged Young's modulus, which is related
to the average shear modulus $G$ (cf.\ \qeq{G})  and  the
average Poisson ratio $\nu$ (cf.\ \qeq{nu})
via

\begin{equation} \label{E}
E= 2G (\nu+1) ,
\end{equation}

and

\begin{equation} \label{7}
\Gamma(\Om) = {\Omega_1}^2 {\Omega_2}^2 +
{\Omega_2}^2 {\Omega_3}^2  +
{\Omega_3}^2 {\Omega_1}^2
\end{equation}

describes the angular dependence.\cite{TS71}

For reference purposes, it is interesting to define pertinent isotropic
properties, such as they would apply to a polycrystalline
sample:\cite{VN93} the bulk modulus
$B$,

\begin{equation}  \label{B}
B = \frac{c_{11}+ 2  c_{12}}{3} ,
\end{equation}

an average shear modulus $G$, defined as the arithmetic mean over the
Voigt- and Reuss-averaged shear moduli

\begin{equation}  \label{G}
G = \frac{1}{2} \left(
\frac{c_{11} -  c_{12}+ 3c_{44} }{2(G+3B)} +
\frac{5(c_{11} - c_{12})c_{44}}{4c_{44} +3 (c_{11} -  c_{12})}
\right) ,
\end{equation}

and an average Poisson ratio $\nu$ as appropriate for an isotropic solid

\begin{equation}  \label{nu}
\nu = \frac{3B-2G}{2(G+3B)} .
\end{equation}

Indentation moduli $\ErOm$ for the three crystal orientations have been
determined numerically by Vlassak and Nix.\cite{VN93,VN94} We evaluate
these moduli for Al and Cu and give the results, denoted as $\Ertheo$ in
Table II. The orientation dependence amounts to only 1 \% for Al, while
for Cu, deviations of up to 10 \% from the isotropic value, $\Eriso$,
show up. In agreement with the orientation dependence of the Young's
modulus, \qfig{aniso}, the (111) plane is the stiffest, and the (100)
plane the softest; the (110) plane is intermediate, and its modulus is
quite close to the isotropic value.

Other authors\cite{TS07}  attempt a different procedure
and  define

\begin{equation} \label{Erapp}
\Erapp(\Om) = \frac{E(\Om)}{1- \nu^2(\Om) } .
\end{equation}

Thus they use the definition of the isotropic indentation modulus
\qeq{Er}, and replace Young's modulus $E$ by the orientation dependent
Young modulus $E(\Om)$. Analogously, $\nu$ is replaced by the Poisson
ratio $\nu(\Om)$, which characterizes the contraction transverse to
$\Om$; for the (100) and (111) orientation, this value is uniquely
defined, while for the (110) direction a Voigt average is applied. The
resulting values of this approximated indentation modulus are displayed
in Table II.
They show a considerably larger anisotropy than the Vlassak-Nix
moduli; the
deviation from the isotropic modulus vary between $-9$ and $+6$ \% for
Al and between $-44$ and $+45$ \% for Cu. Evidently, $\Erapp$ is not a
good approximation to the true theoretical value, $\Ertheo$.

\section{Results} \label{Results}

\subsection{Elastic regime}   \label{s_elastic}

\begin{figure}

\subfigure[]{\includegraphics[width=0.45\textwidth]{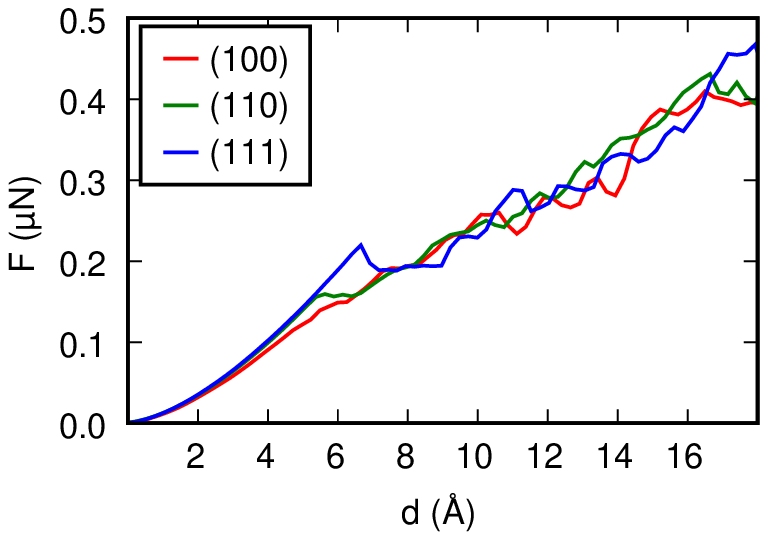}}
\hfill
\subfigure[]{\includegraphics[width=0.45\textwidth]{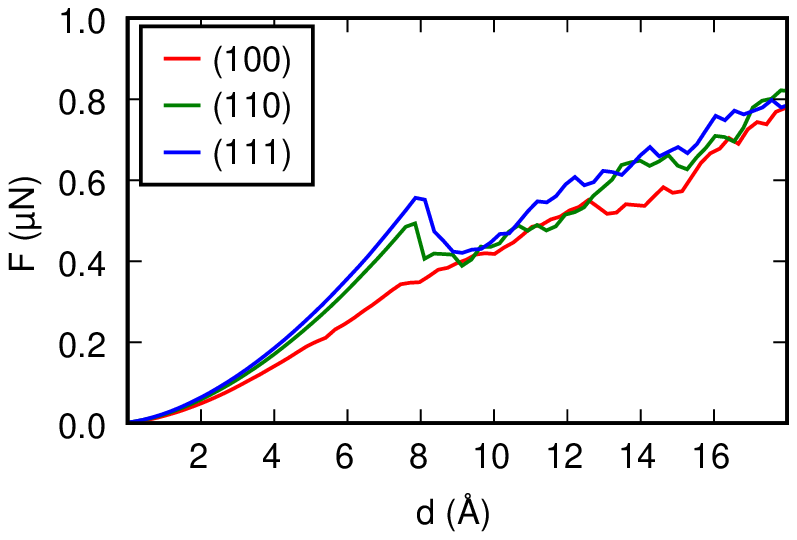}}

\caption{Force $F$ vs displacement $d$ in Al (a) and Cu (b) for the
three surface orientations studied.
 }

\label{f_Fd}
\end{figure}

In \qfig{Fd} we show the force-displacement curves for Al and Cu
indented on the three principal surface planes.
In the elastic regime, which is characterized by the $d^{3/2}$
dependence of the force and terminates with a more or less drastic
force maximum at a depth of around $6-8$ \AA, Cu is stiffer than Al,
corresponding to the larger elastic moduli of Cu.
Both materials show a dependence on the surface orientation,
such that the stiffness is highest for the (111) plane and smallest for
the (100) plane, in agreement with the orientation dependence of the
indentation modulus, cf.\ Table II.
The orientation dependence is most pronounced for Cu, corresponding
to the larger anisotropy $\ani$ of this material.

We fit the elastic part of the force-indentation curve to the Hertzian
law, \qeq{Hertz}, using $\Er$ as a fit parameter. We perform two fits:
(i) over the full elastic part of the $F(d)$ curve up to $d=5$ \AA,
giving a fit parameter $\Erfitlong$; (ii) a fit only the initial part,
$d<2$ \AA, giving $\Erfitshort$. The fit values are included in table
II. We see that the full fit gives consistently larger fitted
indentation moduli than the initial fit, $\Erfitlong> \Erfitshort$.
This is a sign of the onset of nonlinear elasticity: under pressure, a
material shows an increased stiffness; this feature has been termed
\emph{pressure hardening}.\cite{ZLV*04,OLY02} This effect of pressure
hardening has been demonstrated previously by comparing FEM simulations
of the indentation process using linear vs nonlinear elasticity.\cite{ZLV*04}
The fact that the indenter is not exactly a hard sphere, but --
for numerical reasons -- has to be defined with a finite stiffness in form
of the polynomial indenter potential, \qeq{5}, also causes the initial
stiffness to
be systematically smaller than the stiffness at larger penetration depths; this feature
enhances the effect of the pressure hardening.

A comparison of the simulation results with theory hence must be based
on the initial fit values $\Erfitshort$. Table II demonstrates a fairly
good agreement between theory and simulation; the largest deviations
amount to 14 \%. We note that -- with the exception of the Cu (100)
surface -- all fit values are above the theoretical values; this points
at the possibility that even for the small fit regime of $0<d<2$ \AA,
already some departures from linear elasticity may show up. Among other
reasons which may contribute to the deviations we mention
the atomistic nature of the indentation process,
which for the indenter radius of 8 nm is not fully captured by continuum
elasticity, and also
the numerical problem of fitting the
molecular-dynamics data to the Hertzian law -- note that besides the
elastic deformation also a finite offset in the displacement has to be
fitted.

\begin{figure}

\subfigure[]{\includegraphics[width=0.45\textwidth,clip=true]{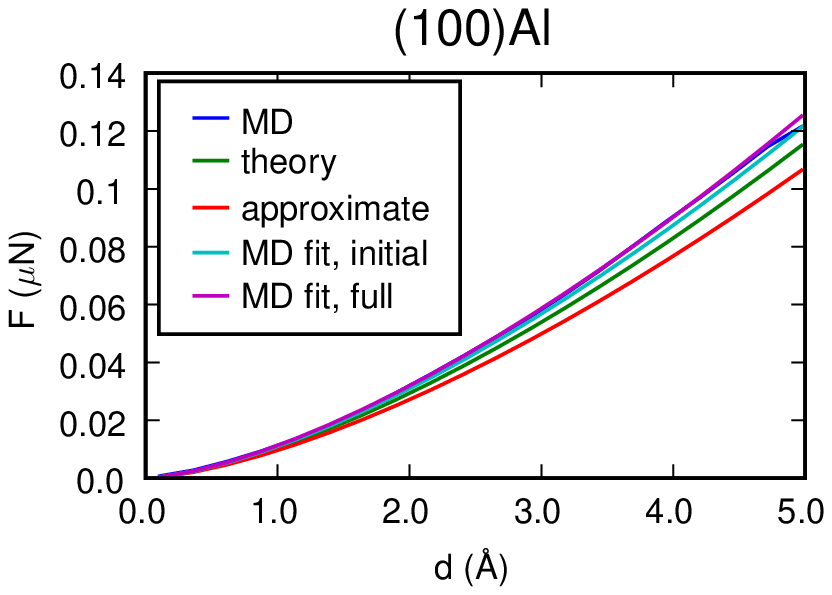}}
\hfill
\subfigure[]{\includegraphics[width=0.45\textwidth,clip=true]{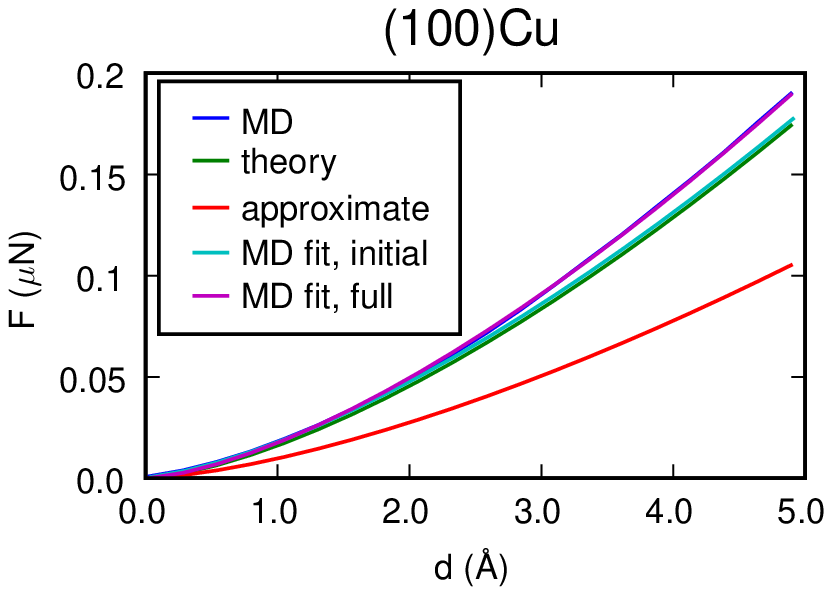}}

\caption{Comparison of the force $F$ vs displacement $d$ data as
obtained by molecular dynamics (MD) with various theories and fits. \\
\emph{theory}: theoretical prediction for anisotropic media by Vlassak
and Nix.\cite{VN93,VN94} \\
\emph{approximate}: \qeq{Erapp}.     \\
\emph{MD fit, initial}: Fit of MD data to Hertz' law, \qeq{Hertz}, for
$d<2$
\AA. \\
\emph{MD fit, full}: Fit of MD data to Hertz' law, \qeq{Hertz}, for the
full elastic region. \\
a) Al (100). b) Cu (100).
 }

\label{f_fit}
\end{figure}

In \qfig{fit} we give a graphical representation of the comparison
between MD simulation results, the fit curves, and the theoretical
prediction for the case of the (100) surfaces. Both for Al and Cu,
the MD data are almost indistinguishable from the full fit. It is
also evident that deviations between the full fit and the initial fit
become sizable only for $d>2$ \AA, as it must be. The approximate
indentation moduli, \qeq{Erapp}, severely underestimate the simulation data,
while the theoretical prediction by Vlassak and Nix gives a fair
representation of the simulation data.

\subsection{Plasticity}   \label{plastic}

The onset of plasticity is visible in the force-displacement curves of
\qfig{Fd} by the pronounced load drop appearing for the (111) and -- to
a lesser extent -- for the (110) surface. In the following we denote the
stress at which this load drop appears as the critical stress and the
pertinent indentation as the critical indentation depth. For the (100)
surface, dislocation nucleation sets in considerably earlier and in a
more continuous way, such that any load drop is blurred out.

These principal differences between the (100) and (111) surfaces and the
abrupt onset of plasticity for the (111) surface can be explained as
follows: (i) the primary glide systems \slip\ are located at quite
oblique angles to the direction of the indentation force acting normally
to the surface; the corresponding Schmid factor is only $\sone =
\sqrt{2/27} = 0.27$. For the (100) surface these glide systems are more
easily activated, since $\szero = 1.5 \sone= 1/\sqrt{6}=0.41$. (ii) When
finally the critical indentation depth has been reached, a considerable
elastic energy has built up due to the high stiffness of this surface.
Then, upon dislocation nucleation, a stronger dislocation avalanche and
consequently a larger plastic displacement jump are achieved. The (110)
surface shows a somewhat more complex indentation behaviour; while the
pertinent Schmid factor is identical to the (100) surface, the onset of
plasticity rather follows the (111) surface. We assume that this is due
to the complex stress distribution found below this surface; in
particular, the maximum shear stress is not below the indenter along the
indentation axis but rather shifted to the side; we also observe more
than one primary nucleation site for dislocations. Note also that it has
been shown\cite{TSM07,TM08} that -- in single-crystalline Cu --
non-Schmid factors may be important in describing dislocation nucleation
for uniaxial tension in [110] direction, but not in [111] or in [100]
direction.

The influence of the surface orientation on the force-displacement
curves increases in the elastic regime until the critical stress is
reached and the load drop appears. At this critical indentation depth,
the orientation dependence is maximum, i.e., the forces exerted on the
three principal surfaces for identical indentation vary most. Due to the
stronger anisotropy of Cu, the orientation dependence is stronger for
this material than for Al. Note, however, that also in Al, which is
almost isotropic, a distinct orientation dependence is seen around the
critical indentation depth; this effect is not due to the elastic
anisotropy but rather to the processes of dislocation nucleation, see
the discussion above. After the load drop, in both materials the
force-displacement curves show almost no orientation dependence, apart
form fluctuations. and increase linearly with depth.

This feature is better discussed with the help of the contact
pressure, that is the force divided by the projected contact area. Note
that the measurement of this area in an atomistic simulation is not
without ambiguities, see the Appendix.
The
pertinent data are displayed in \qfig{pd}. Initially, before the load
drop, the pressure curves increase in agreement with the
$\sqrt{d}$-dependence of the Hertzian theory, \qeq{p}. The pressure for
the (111) and the (110) surface increase quite similarly, in agreement
with the quite similar values of the indentation moduli for these two
orientations, cf.\ Table II. The pressure on the (100)
surface, however, starts deviating quite early from that of the other
surfaces, and is considerably lower; this is in agreement with the above
discussion of the earlier dislocation nucleation occurring under this
surface. The load drop shows up in a more pronounced way in this plot,
since the projected areas increase considerably during and after the
nucleation of plasticity.

After the end of the load drop, the contact
pressure reaches a rather constant value;   this defines the hardness of
the material. Thus \qfig{pd} demonstrates that the surface orientation
loses its influence on the hardness as soon as the plastic regime has
been entered. The hardness of Al is measured in the simulation to be
around 6 GPa, and that of Cu around 12 GPa; these values correspond
quite well to the simulational data obtained in \qref{TS07}.
Experimental data as obtained by nanoindentation in single-crystalline
Cu
(100) give only slightly smaller values, around 6 GPa,\cite{GK01} while
measurements on (ultra-fine grained)  poly-crystalline specimens obtain
smaller hardness values, around 2 GPa.\cite{DBG05}

\section{Conclusions}

We performed simulations of nanoindentation into an elastically
isotropic
metal, Al, and an elastically anisotropic metal, Cu. The three principal
surfaces -- the (100), (110) and (111) surface -- were studied. We
found:

\begin{enumerate}

\item The elastic deformation is fairly well described by linear
elasticity theory, as tabulated by Vlassak and Nix. With increasing
indentation, the effect of pressure hardening is visible.

\item Around the critical indentation depth, just before the onset of
dislocation nucleation, the indentation forces show the strongest
dependence on surface orientation. This dependence is almost equally
strong in the elastic isotropic Al as in Cu. It is not an elastic effect
but due to the fact that dislocations nucleate at different  global
stresses due to the different orientation of the slip planes.

\item After the load drop, the contact pressure inside the material
stays roughly constant and assumes the same value independently of the
surface orientation. This demonstrates that the indentation hardness is a
well defined concept even when the critical indentation depth has been
only slightly exceeded.

\end{enumerate}

\begin{acknowledgments}

The authors acknowledge financial support by the
Deutsche Forschungsgemeinschaft via the Graduiertenkolleg 814, and a
generous grant of computation time from the ITWM, Kaiserslautern.

\end{acknowledgments}

\appendix
\section{How to determine the contact area?}

The choice of the contact area is crucial for determining the contact
pressure, but unfortunately not unique. On the macroscopic length scale
at least two conceptionally different concepts exist: the Meyer and the
Brinell contact areas, \cite{Fis04} which represent the convex contact
surface and the projection onto the initial surface plane, respectively.

\begin{figure}

\includegraphics[width=0.45\textwidth]{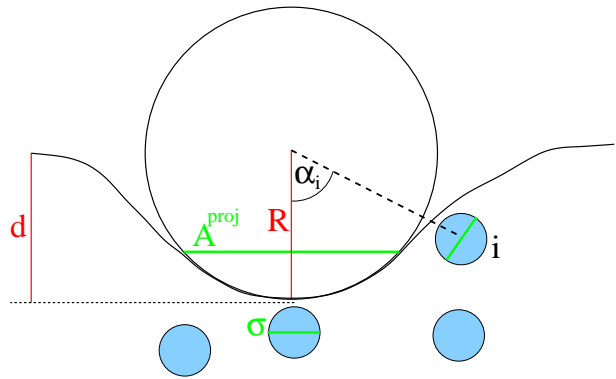}

\caption{ Schematics of nanoindentation by a sphere of radius $R$. The
indentation depth $d$ is measured with respect to the initial surface
plane. The relevant area $\Apro$ is the contact area of the indenter
with the substrate projected into the initial surface plane.
$\alpha_i$ is the angle
between the indentation direction  and the  vector joining the centre of
the indenting sphere with atom $i$.
 }

\label{f_scheme}
\end{figure}

For the purposes of nanoindentation, and in agreement with Hertz, the
contact pressure is the substrate response in normal direction and
therefore we have to choose the projected area $\Aatom$. On the atomistic
level the obvious choice to measure this area is to sum up the
(projected) areas of all atoms $i$ which are in contact with the
indenter, cf.\ \qfig{scheme}:

\begin{equation}      \label{area1}
   \Aatom =\pi\sigma^2\sum_{i\in\mathrm{contact}}
\cos \alpha_i .
\end{equation}

\begin{figure}

\includegraphics[width=0.45\textwidth]{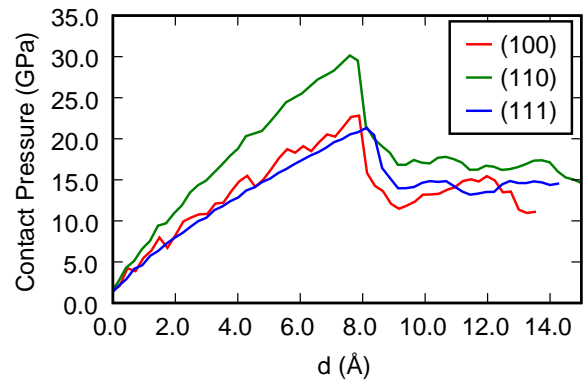}

\caption{Contact pressure  vs displacement $d$ in Cu for the
three surface orientations studied, as measured with a too simplistic
definition of the indenter contact area, \qeq{area1}.
 }

\label{f_App}
\end{figure}

Here, $\sigma$ is an `atom radius', and $\alpha_i$ is the angle between
the indentation direction and the vector joining the centre of the
indenting sphere with atom $i$. For Cu we choose $\sigma=1.35$ \AA.
Unfortunately, this definition leads to serious problems, since $\Aatom$
describes a non-connected area; indeed, between the atoms there appear
`holes'. This becomes an important problem in particular in the elastic
regime, where due to the elastic deformation, these `holes' increase in
size with increasing deformation, leading to systematic errors: the area
is underestimated, and the pressure is overestimated. A further
systematic effect shows up due to the different areal densities of
surface planes: For atomically rough surfaces, such as the (110), the
area is underestimated in comparison to dense surfaces, such as the
(111); this effect leads to a distortion of hardness determinations from
simulation data for different surfaces. \qfig{App} demonstrates these
effects: besides a general overestimation of the contact area, the order
of the stiffness for the (111) and the (110) surfaces is exchanged.

In the present paper, we therefore have chosen an elliptic contact area,
which is connected and conserves the right order of the pressures:

\begin{equation}   \label{area2}
   A^\mathrm{elliptic} =\frac{\pi}{4}
(\xmax - \xmin) (\ymax - \ymin) .
\end{equation}

\begin{figure}

\subfigure[]{\includegraphics[width=0.45\textwidth]{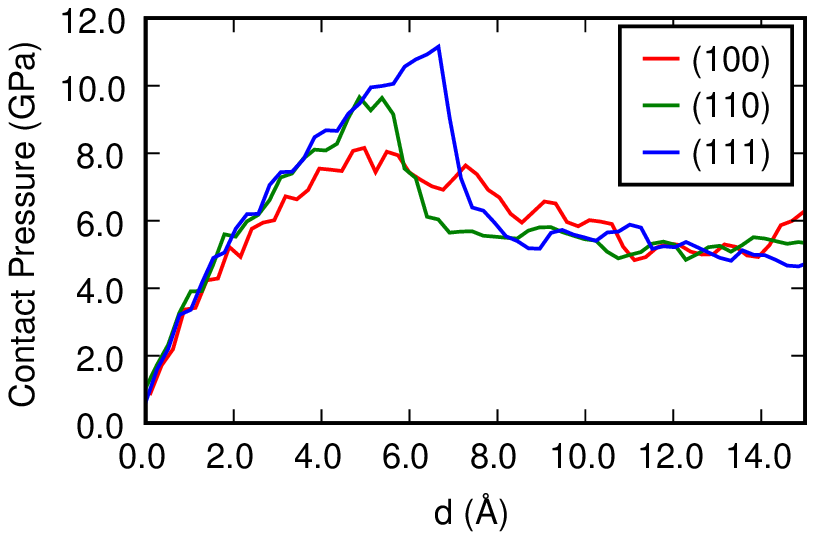}}
\hfill
\subfigure[]{\includegraphics[width=0.45\textwidth]{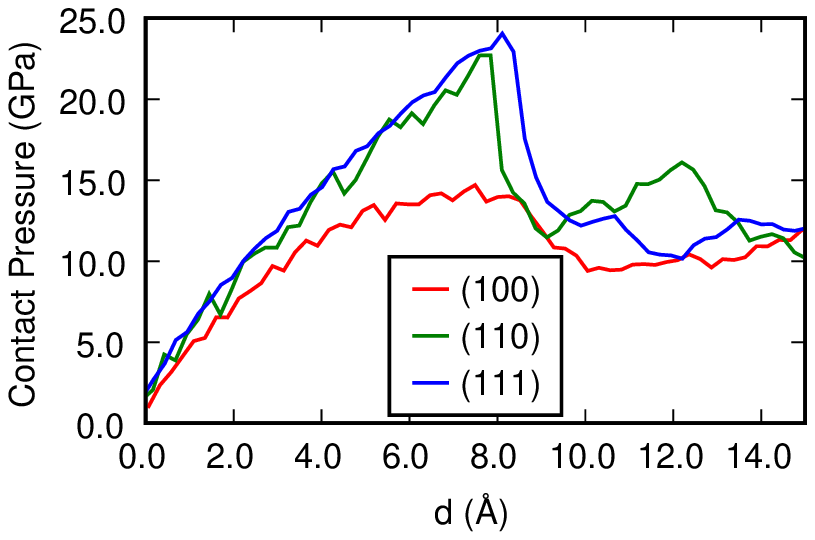}}

\caption{Contact pressure  vs displacement $d$ in Al (a) and Cu (b)  for the
three surface orientations studied.
 }

\label{f_pd}
\end{figure}

Here, $x$ and $y$ measure the coordinates of the contact atoms projected
into the initial surface plane. This set of coordinates describes a
curved contour line, which is approximated by an ellipse. In
\qeq{area2}, $\xmax - \xmin$ and $\ymax - \ymin$ denote the major and
minor diameters of this ellipse, respectively. This definition leads
initially, i.e., when the indenter has contact with few atoms only, to
an increased noise in the determination of the area and hence the
contact pressure, cf.\ \qfig{pd}. For larger indentations, however, it
is more reliable.


\bibliography{c:/D/bib/base/string,c:/D/bib/base/all,c:/D/bib/base/publ}


\begin{table}

\caption{ Elastic properties of Al and Cu as described by the potentials
by Zhou et al.\cite{ZWJ*01} and by Mishin et al.\cite{MMP*01}.
Elastic
constants $c_{ij}$, elastic anisotropy $\ani$, \qeq{ani}, bulk modulus
$B$,  average shear modulus $G$, and average Poisson ratio $\nu$.
 }

\vspace{5ex}
\begin{tabular}{l|lll|l|lll}
   &  $c_{11}$ (GPa) &  $c_{12}$ (GPa)  &  $c_{44}$ (GPa)  & $\ani$   &  $B$ (GPa)  &  $G$ (GPa)  &  $\nu$ \\
\hline
Al & 114   & 61.5  & 31.6 & 1.20 & 79.0  & 29.3 & 0.33 \\
Cu & 169.9 & 122.6 & 76.2 & 3.22 & 138.4 & 47.8 & 0.35
\end{tabular}

\label{T1} \end{table}

\begin{table}

\caption{ Orientation dependent indentation moduli. \\
$\Ertheo$: after
Vlassak and Nix\cite{VN93,VN94}. \\
$\Erapp$: obtained according to the
simplified recipe, \qeq{Erapp}. \\
$\Erfitshort$: from a fit to our MD
results over $d<2$ \AA. \\
$\Erfitlong$: from a fit to our MD results over
the full elastic region. \\
The isotropic indentation modulus $\Eriso$
as calculated from the isotropic constants, Eqs.~\queq{B} -- \queq{nu},
appropriate for a polycrystal, is given as a reference.
 }

\vspace{5ex}
\begin{tabular}{l|lll|lll|}
   &    \multicolumn{3}{c|}{Al}    &    \multicolumn{3}{c|}{Cu} \\
\cline{2-4} \cline{5-7}
   &  (100) & (110) & (111) &  (100) & (110) & (111) \\
\hline
$\Ertheo$ (GPa)       & 87.0  & 88.4 & 88.9  & 135.0 & 148.0 & 151.9 \\
$\Erapp$ (GPa)        & 80.5  & 90.3 & 93.4  & 81.4  & 179   & 211   \\ \hline
$\Erfitshort$ (GPa)   & 92    & 96   & 101   & 134   & 155   & 171   \\
$\Erfitlong$ (GPa)    & 95    & 97   & 106   & 147   & 176   & 192   \\ \hline
$\Eriso$ (GPa)     &\multicolumn{3}{c|}{88.2} &\multicolumn{3}{c|}{145.9}
\end{tabular}

\label{T2} \end{table}


\clearpage

\end{document}